\documentclass[aps,prl,twocolumn,superscriptaddress,showpacs]{revtex4}

\usepackage[dvips]{graphicx}
\usepackage{latexsym}
\usepackage{bm}

\begin{document}

\title{
Direct Numerical Confirmation of Pinning Induced Sign Change in 
Superconducting Hall Effect in Type-II Superconductors
}

\author{
Noriyuki Nakai
}
\email[]{nakai.noriyuki@jaea.go.jp}
\affiliation{
CCSE, Japan Atomic Energy Agency, 6-9-3 Higashi-Ueno, Taito-ku, Tokyo 110-0015, Japan
}
\affiliation{
CREST(JST), 4-1-8 Honcho, Kawaguchi, Saitama 332-0012, Japan
}

\author{Nobuhiko Hayashi
}
\affiliation{
CREST(JST), 4-1-8 Honcho, Kawaguchi, Saitama 332-0012, Japan
}
\affiliation{
Nanoscience and Nanotechnology Research Center (N2RC), Osaka Prefecture University,
1-2 Gakuen-cho, 
Sakai 599-8570, Japan
%
%
}

\author{Masahiko Machida
}
\affiliation{
CCSE, Japan Atomic Energy Agency, 6-9-3 Higashi-Ueno, Taito-ku, Tokyo 110-0015, Japan
}
\affiliation{
CREST(JST), 4-1-8 Honcho, Kawaguchi, Saitama 332-0012, Japan
}
\date{\today}

\begin{abstract}
Using the time-dependent Ginzburg-Landau equation with the complex relaxation time and the Maxwell equation, 
we systematically examine transverse motion of vortex dynamics in the presence of pinning disorders. 
Consequently, in a plastic flow phase in which moving and pinned vortices coexist, we find 
that the Hall voltage generally changes its sign. The origin of the sign change is ascribed to 
a fact that moving vortices are strongly
drifted by circular current of pinned vortices and the enforced
transverse moving direction becomes opposite to that by transport current.
This suggests that the Hall sign change is a behavior common in all disordered type-II superconductors.
\end{abstract}


\maketitle

\begin{figure*}[tb]
\includegraphics[width=17cm,keepaspectratio]{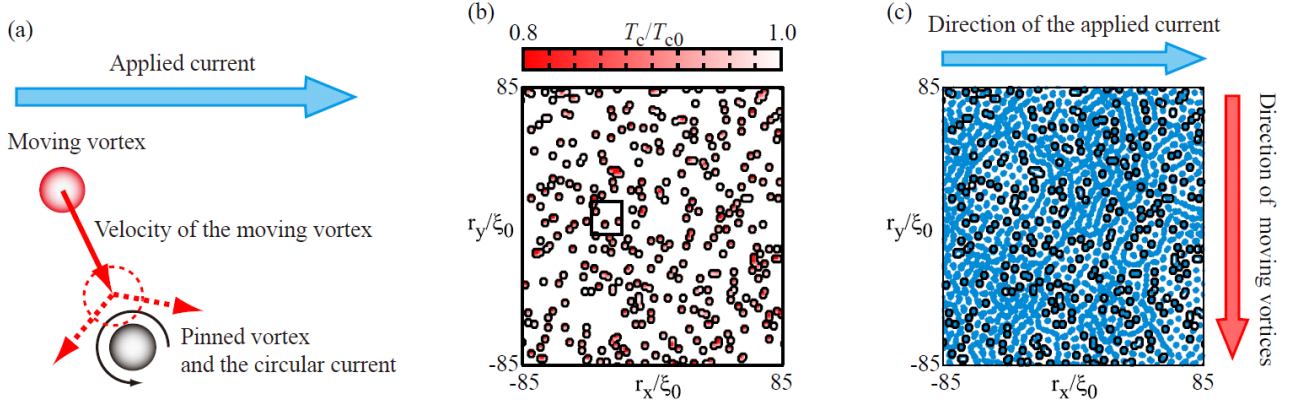}
\caption{
(Color online)
(a) A schematic figure of a moving single vortex driven by an applied current
in the case of a complex TDGL relaxation rate in the presence of the pinned vortex.
(b) The contour map of the local transition temperature $T_{\rm c}({\bm r})$. 
The closed curves (black) show the positions of pinning sites,
and the shading color indicates degree of the $T_{\rm c}$ suppression.
(c) The superimposed snapshots of moving vortices
in the time interval $4 \times 10^5 \le t/t_0 \le 6 \times 10^5$.
The vortices are visualized at each instant
by the contour plot (blue) of the order parameter at $\Delta/\Delta_0=0.2$.
The positions of pinning sites are marked by black curves.
$T/T_{\rm c0}=0.69$, $H_a/H_0=0.2$, and $j_x/j_0= 5 \times 10^{-5}$.
}
\label{fig1}
\end{figure*}
\begin{figure*}[htb]
\begin{center}
\includegraphics[width=17cm,keepaspectratio]{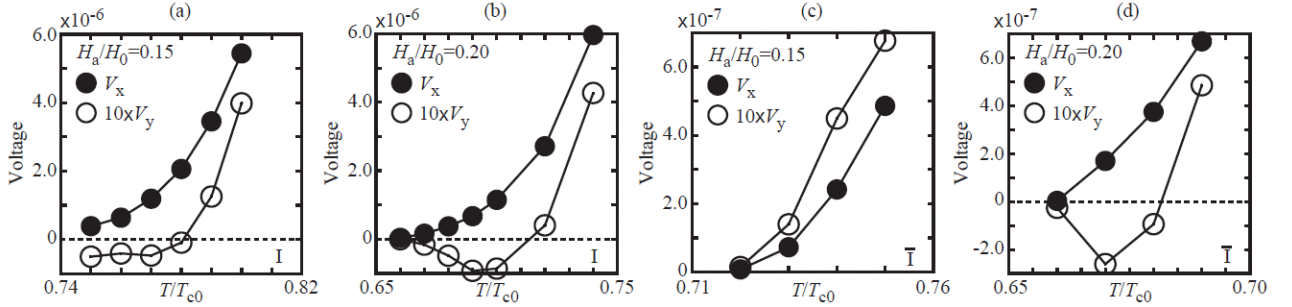}
\caption{
Temperature dependences of the longitudinal ($V_x$) and Hall ($V_y$) voltages
in units of $A_0 \xi_0/t_0$.
The filled (open) circles indicate the longitudinal (Hall) voltage.
Note that the Hall voltage is plotted in the ten times larger scale.
The voltages are calculated from
the electric field averaged over
$-85 \le r_{x(y)}/\xi_0 \le 85$ and $4 \times 10^5 \le t/t_0 \le 6 \times 10^5$.
The applied current is $j_x/j_0= 5 \times 10^{-5}$.
The applied field is $H_a/H_0=0.15$ for (a) and (c), and $H_a/H_0=0.2$ for (b) and (d).
The results in (a) and (b) are obtained for
the distribution of pinning sites shown in Fig.~1(b),
while the same distribution is used but whose bottom is inverted into top for moving vortices 
to obtain the results (c) and (d) (see text).
}
\label{fig2}
\end{center}
\end{figure*}
\begin{figure*}[htb]
\begin{center}
\includegraphics[width=17cm,keepaspectratio]{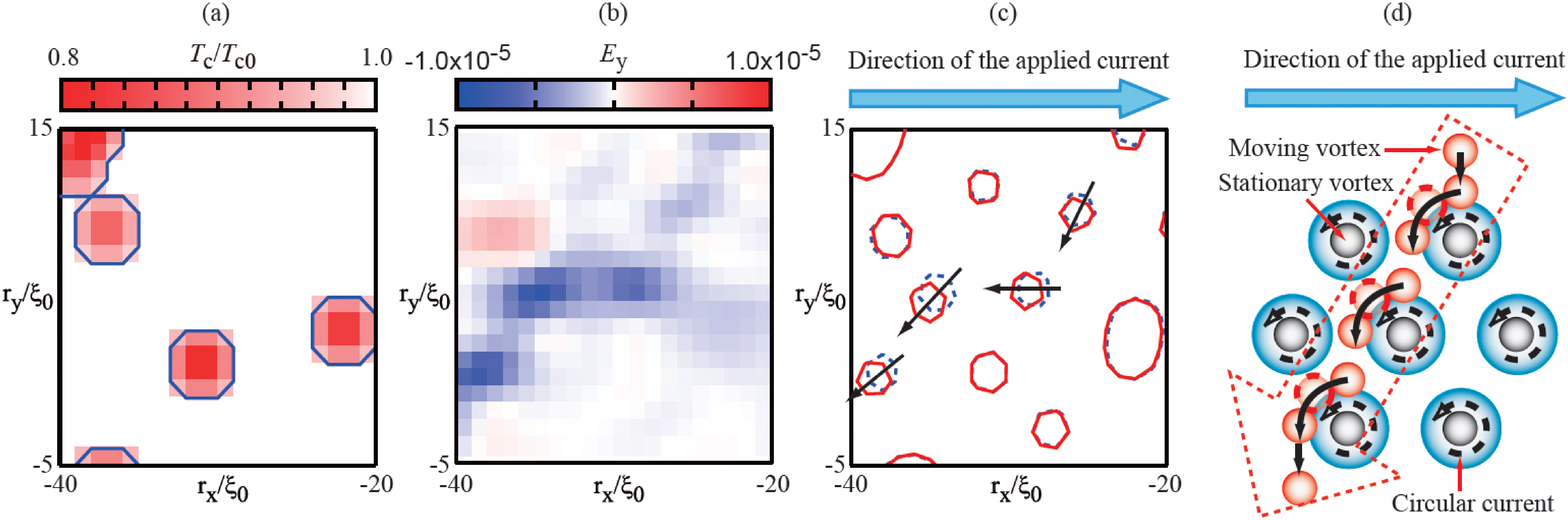}
\caption{
(Color online) 
(a) The distribution of pinning sites in the focused area, whose location 
is marked by the small box in Fig.~1(b).
(b)The transverse electric field in the area
for $T/T_{\rm c}=0.69$, $H_a/H_0=0.2$, and $j_x/j_0=5 \times 10^{-5}$.
The data is averaged over $5.8 \times 10^5 \le t/t_0 \le 5.84 \times 10^5$.
(c)The vortex positions at $t/t_0=5.8 \times 10^5 $ and $5.84 \times 10^5 $ through 
the contour plot of the order parameter amplitude. 
The arrows indicate the moving directions
(d)The schematic representation of moving vortices around stationary vortices.
}
\label{fig3}
\end{center}
\end{figure*}


Since the discovery of cuprate high-$T_{\rm c}$ superconductors, much attention has been 
devoted to vortex dynamics in not only superconductors but also various superfluids from 
liquid Helium to atomic gas.
In particular, vortex pinning dynamics under disorders inevitable 
in superconducting materials is a central issue of vortex physics because of 
deep relations to its industrial applications. 
In this paper, we numerically examine the vortex pinning dynamics and 
give a clear explanation to a controversial topic in vortex physics, i.e., sign change in 
the superconducting Hall effect 
\cite{hagen,matsuda,nagaoka,kokubo,ping-ao}.

The equation of motion for a moving vortex in superfluids
has been highly controversial \cite{ping-ao}.
The heart of the problem is
the non-dissipative transverse force
(or the vortex velocity part of Magnus force) \cite{volovik96,ao-kopnin-prlcomrep,hall-wexler-prlcomrep,zhu},
which brings about the superconducting Hall effect. 
In BCS superfluids, the transverse component is generally subtle, and therefore, various theoretical proposals 
remain unexamined fully. 
Among those theories, one of the most controversial struggles is whether pinning or disorder can be an origin of 
the sign change in the superconducting Hall effect \cite{vinokur-1993,wang,ao-1998,kopnin-1999,ikeda,b-y-zhu,ghenim}. 
If it is true, it then indicates 
that the sign reversal is not limited in particular superconductors, but universal for all disordered type-II superconductors.
In this paper, we clarify that the idea is really true by numerically solving the time-dependent Ginzburg-Landau (TDGL) equation 
with complex relaxation time and the Maxwell equation.  
This is a direct and clear confirmation of the pinning induced sign change without 
simplification and modeling.

The vortex dynamical phases under disorders are roughly classified into two types, i.e., 
plastic and collective flow phases \cite{olson}.
In the former phase, moving and pinned stationary vortices coexist.
Then, it is found that moving vortices are strongly drifted by the circular current of pinned vortices and 
its inducing transverse moving-direction can become opposite to that by transport current.
Such a feature is schematically displayed in Fig.~1(a).
Thus, the averaged transverse moving direction in the plastic flow phase can be different 
from that in the collective one. 
This is an origin of the Hall sign change confirmed in this paper.


Let us present the system setup to confirm the pinning induced sign change.
We prepare a two dimensional system in the $xy$ plane.
The external magnetic field $H_a$ is applied perpendicular to the plane.
To simulate vortex dynamics, we numerically solve the 
the TDGL equation coupled with the Maxwell one
written as \cite{kato,machida,crabtree,nakai08}
\begin{eqnarray}
\frac{\partial \Delta}
{\partial t}
&=&
-\frac{c^2}{48\pi\kappa^2\xi_0^2\sigma\Gamma}
\biggl[
\biggl\{
\left|
\frac{\Delta}{\Delta_0}
\right|^2
-\left(
1-\frac{T}{T_{\rm c}}
\right)
\biggr\}
\Delta
\nonumber \\
& & { } +
\xi_0^2
\left(
\frac{\bm{\nabla}}{ i}
-
\frac{2e}{\hbar c}{\bm A}
\right)^2
\Delta
\biggr],
\label{gleq} \\
%
\frac{\sigma}{c}
\frac{\partial {\bm A}}
{\partial t}
&=&
\frac{\hbar c^2}{8\pi e \kappa^2\xi_0^2\Delta_0^2}
\biggl[
\Delta
\left(
-
\frac{\bm \nabla}{ i}
-
\frac{2e}{\hbar c} {\bm A}
\right)\Delta^{\ast}
\nonumber \\
& & { }
+
\Delta^{\ast}
\left(
\frac{\bm \nabla}{ i}
-
\frac{2e}{\hbar c} {\bm A}
\right)
\Delta
\bigg]
-\frac{c}{4\pi}
{\rm rot}
{\bm H}.
\label{mx}
\end{eqnarray} 
Here, we introduce local suppressions of the transition temperature
$T_{\rm c} ({\bm r})$, which act as vortex pinning sites.
The order parameter $\Delta$ is normalized by its mean field value at the zero temperature 
without the magnetic field, 
$\Delta_0$, and time $t$, vector potential ${\bm A}$, and magnetic field ${\bm H}$
are done by 
$t_0=4\pi\kappa^2\xi_0^2\sigma/c^2$, $A_0=\phi_0/(2\pi\xi_0)$, and $H_0=\phi_0/(2\pi\xi_0^2)$, 
respectively, where $\xi_0$, $\kappa$, $\sigma$, $c$, and $\phi_0(=2\pi\hbar c/(2e))$ are the zero-temperature coherence length, 
the Ginzburg-Landau parameter, the normal-state longitudinal conductance, the light velocity, and 
the flux quantum, respectively. 
To keep the gauge invariance of Eqs.~(1) and (2) on numerical grids, we use the 
link variable 
$U_{\mu}^{ij}
={\rm exp}
\left[
-{ i}\int_{r_i}^{r_j}
({A_\mu}/{A_0})
{d\mu}/{\xi_0}
\right]$,
where $\mu$ stands for $x$ or $y$ \cite{kato,machida,crabtree,nakai08}.
The magnetic field ${\bm H}$ is given 
by the Stokes' theorem
$
\int_S ({\bm H}/H_0)\cdot {\bm n}dS/\xi_0^2
=\int_c
({\bm A}/A_0)
\cdot d\,{\bm l}/\xi_0
$, and the electric field is calculated by
$E_\mu=-(A_0/t_0)\int_{\bar S} \partial (A_\mu/A_0)/\partial (t/t_0) d^2r /{\bar S}$,
where ${\bar S}$ is the unit plaquette surrounded by link variables.
We evaluate the longitudinal and the Hall voltage from $E_\mu$.
In order to concentrate on the vortex contribution to the Hall voltage, we 
neglect the normal-state Hall conductivity in Eq.\ (\ref{mx}) for clarity.
Instead, the dimensionless relaxation rate $\Gamma$ in Eq.\ (\ref{gleq}) 
is set to a complex number. According to Ref.\ \cite{dorsey},
$1/\Gamma$ is related to the forces acting on each moving vortex.
If we set $1/\Gamma$ pure real,
the transverse force and the resulting Hall voltage are zero.
On the other hand, a finite imaginary part of $1/\Gamma$ brings about
a transverse force, and the sign of the imaginary part controls 
the sign of the Hall effect \cite{dorsey,fukuyama,kopnin-TDGL,aronov,matsuda}.
{\it We keep the imaginary part positive and never change its value throughout this study},
i.e., the transverse force is always positive and unchanged.
Under the condition of the fixed complex relaxation rate,
we find that the Hall voltage amplitude diminishes and a sign reversal occurs in the plastic flow regime
owing to moving vortices affected by the circular current around pinned vortices.

In the present simulation, the system size is $200\xi_0 \times 200\xi_0$,
which is discretized by the square grid whose unit dimension is
$\xi_0\times \xi_0$. The external current is applied along the $x$ direction,
and a periodic boundary condition is imposed in this direction to eliminate 
edge boundary effects on the vortex motion. 
The remaining boundary edge perpendicular to the $y$ direction relevant to 
vortex entry and escape is modeled 
as an interface between a superconductor and a normal metal.
Around this interface, $T_{\rm c}({\bm r})$ is set $T_{\rm c}/T_{\rm c0}=0.1 r/\xi_0$, 
where $r$ is the distance from the interface
and $T_{\rm c0}$ is the bulk value of $T_{\rm c}$.
The number of vortex pinning centers is 500 inside 
the present system $200\xi_0 \times 200\xi_0$.
The size of each pinning center is $2 \xi_0 \times 2 \xi_0$, inside which $T_{\rm c}({\bm r})$ is randomly
suppressed in the range $0.8\le T_{\rm c}/T_{\rm c0} \le 1$.
The locations of pinning centers are randomly distributed, e.g., 
as shown in Fig.~1(b).

In the TDGL dynamical simulation, we prepare an initial state in the absence of both the applied current and external field, 
and then start to apply a current $j_x$ and a target external field $H_a$ at $t=0$.
The applied current density is always set as $j_x/j_0=5 \times 10^{-5}$, where $j_0(=\phi_0/(2\pi\xi_0^3))$ is 
the depairing current. The GL parameter is $\kappa=2.83$, and the minimal time step is $3 \times 10^{-3} t_0$.
We fix $1/\Gamma=1+0.3{ i}$, which leads to a substantial ratio of the Hall voltage $V_y$
to the longitudinal one $V_x$, i.e., $V_y/V_x \sim 0.2$, in the uniform current under no pinnings.
Such a large imaginary value can give a striking contrast in the sign change of the Hall 
voltage.  
To avoid counting an interface influence on the voltage, e.g., an effect of the diamagnetic current,
we take an average of the electric field 
within the region $-85 \le r_{x(y)}/\xi_0 \le 85$.
The time average of the Hall voltage is taken over the time interval 
$4 \times 10^5 \le t/t_0 \le 6 \times 10^5$,
during which vortex motions are steady.



Let us present simulation results.
First, the temperature ($T$) dependences of the longitudinal and Hall voltages
under the applied fields $H_a/H_0=0.15$ and $0.2$ are displayed in Figs.~2(a) and 2(b),
respectively.
The longitudinal voltage $V_x$ monotonically decreases with decreasing $T$ in both cases.
Although the Hall voltage $V_y$ exhibits non-monotonic behavior, 
both the signs become negative in the region of the small longitudinal voltage $V_x$.
Here, one might imagine that this negative transverse voltage occurs 
because of guided vortex flow lines  \cite{kopnin-1999} formed accidentally by clustering of pinned sites.
However, it is not always the case.
In order to confirm it, we repeat the simulation by just reversing 
the magnetic field direction only. 
While the sign reversal disappears for the field $H_a/H_0=0.15$ (Fig.~2(c)),
it is kept for another field $H_a/H_0=0.2$ (Fig.~2(d)).
This suggests that there must be an intrinsic origin of the sign reversal 
beyond the guiding effects.
We further perform simulations (not shown)
under other random vortex-pinning distributions (II)--(IV)
and repeat the simulations with the field direction reversal ($\overline{\rm II}$)--($\overline{\rm IV}$).
These results are summarized in Table~I.
The guiding vortex flow state is found to indeed cause the sign change 
under the distributions II and III, because the Hall sign does not change on
the field inversion cases in both II and III.
However, the sign reversal is kept on the field direction reversal for 
the other distributions I and IV.
From these results, it is deduced that the intrinsic sign reversal effect  
always coexists with the guiding one in the plastic flow phase.
One naturally expects that the intrinsic mechanism dominates in sufficiently large 
samples. 
The following analysis attributes the intrinsic mechanism to an effect of the circular currents around pinned vortices.

\begin{table}[tb]
\caption{Yes/No table in terms of the sign reversal of the Hall voltage
for different random pinning distributions (I)--(IV)
and their field inversion cases ($\overline{\rm I}$)--($\overline{\rm IV}$).
``Y(N)'' means that the sign reversal is observable (or not).
}
\begin{ruledtabular}
\begin{tabular}{ccccccccccc}
& & I & $\overline{\rm I}$ &
II & $\overline{\rm II}$ &
III & $\overline{\rm III}$ &
IV & $\overline{\rm IV}$ \\ \hline
\\
& $H_a/H_0=0.15$ & 
Y &
N &
Y &
N &
N &
Y &
Y &
Y \\
& $H_a/H_0=0.2$ & 
Y &
Y &
Y &
N &
N &
Y &
Y &
Y \\
\end{tabular}
\end{ruledtabular}
\end{table}

Let us focus on detailed vortex dynamics around vortex-pinning sites.
Figure 3(a) is an enlarged figure of a small area whose location is 
marked in Fig.~1(b).
The distribution of the transverse electric field averaged over $5.8 \times 10^5 \le t/t_0 \le 5.84  \times 10^5$ 
is shown in Fig.~3(b), where the sign of the transverse field is negative near the vortex pining sites. 
In addition,
by monitoring the vortex position at $ 5.8 \times 10^5 t_0$ and $5.84  \times 10^5 t_0$ in the displayed area,
the focused vortices are found to flow against the applied transport current [Fig.~3(c)]. 
These observed results signify the following. 
The pinned vortex has a circular current flow around its core.
The flow direction is opposite to that of the transport current in 
its top half as schematically shown in Fig.~3(d),
where locally the Lorentz force (or the superfluid velocity part of Magnus force)
is directed from bottom to top.
Then, the moving vortex penetrating into the circular-current flow range 
is drifted into the opposite direction to the transport current.
This is because the positive imaginary part of $\Gamma$ fixed in this paper always drives 
the moving vortex into the downstream side of the current flow as usual. 
Thus, the moving vortex exhibits a motion directed opposite to the transport current.

Finally, let us discuss the present results through a comparison with experiments. 
The present calculations have revealed that 
when the vortex dynamics change from the collective flow to the plastic flow phase 
the moving vortex frequently reverses its transverse moving direction. 
This directional reversal principally requires the plastic flow phase as a vortex dynamical phase.
In other words, this effect is universal for all type-II superconductors as long as disorders or pinning 
sites enough to keep the plastic flow phase are introduced inside the sample.
In high-$T_{\rm c}$ cuprate superconductors, double sign change in addition to single one have been frequently
observed depending on the sample \cite{double-sign-change}.
These experimental results can be explained on the basis of the present result
as follows.
When the current carrying phase changes from the normal to the flux flow phase, the first 
sign change occurs. This can be interpreted by an idea that there is a difference between 
the Hall effect in the normal phase and  
the fluctuation Hall effect near the superconducting transition
 via the relaxation of the order parameter \cite{matsuda,dorsey,fukuyama,kopnin-TDGL,aronov}.
This is a microscopic sign change mechanism depending on the electronic structure. 
On the other hand, it has been observed that
the final reversals strongly depend on the sample quality or the rate of artificial damage  
to enrich pinning centers \cite{artificial damage}.
Thus, the final sign change is attributed to 
the pinning induced one as confirmed by the present simulation.
Moreover, such a sign change is also well-known to be sensitively dependent on the sample quality 
in conventional type-II superconductors.


In conclusion, we performed TDGL simulations with  the complex relaxation time
to confirm the pinning induced sign reversal of the superconducting Hall effect.
Consequently, the simulation revealed that the sign change can occur 
when the current carrying state enters the plastic 
flow phase from the collective flux flow one. 
Moreover, the detailed analysis on the vortex motions 
successfully explained that, when the circular current of the pinned vortex 
strongly drifts the moving vortex in the plastic flow phase, the moving vortex 
feels the transverse force causing the Hall sign change.
These results suggest that the Hall sign change is an indicator of vortex dynamical phases 
in disordered type-II superconductors.


%
\bibliography{basename of .bib file}

\begin{thebibliography}{99}
\bibitem{artificial damage}
See Table I in Ref. \cite{hagen}.
Two limiting cases of the dirty and clean does not show the sign change of the Hall voltage
in V and Nb.
\end{thebibliography}

\end{document}